%% file: main.tex
\documentclass[conference]{IEEEtran}
\IEEEoverridecommandlockouts

\newif\ifblind
\blindfalse         

\newif\ifarxiv
\arxivtrue          

\input{settings/packages}

\input{settings/acronyms}

\begin{document}

\title{Accelerating Precise End-to-End Simulation: Latency-Sensitive Many-core System Modeling
\ifblind
\else
\thanks{
This work has received funding from the Swiss State Secretariat for Education, Research, and Innovation (SERI) under the SwissChips initiative.
\ifarxiv\protect\\
\textcopyright~This work has been submitted to the IEEE for possible publication. Copyright may be transferred without notice, after which this version may no longer be accessible.
\fi
}
\fi
}

\input{settings/authorship}

\maketitle

\begin{abstract}
Modern large language model workloads put increasing demands on parallel compute capability and on-chip memory capacity, while also stressing fine-grained data movement and synchronization.
These trends motivate exploring and designing many-core accelerators with tightly coupled scratchpad memory (SPM) for scalable compute and predictable, explicitly managed data access.
However, this architectural shift raises two challenges: cycle-accurate register-transfer level (RTL) simulation becomes prohibitively slow as system complexity grows, and performance estimation requires precise modeling of latency-sensitive interconnect behavior.
This paper presents a fast yet accurate end-to-end modeling approach for latency-sensitive many-core architectures, targeting large-scale instances such as TeraNoC with 1024 cores and a 4\,MiB globally shared L1 SPM.
The approach captures timing behavior of latency-sensitive SPM accesses across multiple interconnect scales, while abstracting non-essential hardware details.
Across diverse benchmarks, the model tracks a cycle-accurate RTL golden model with errors below 7\%, while delivering up to 115× faster simulation.
The framework also provides detailed profiling across processing elements and interconnect, enabling efficient end-to-end software development and hardware design exploration.
Two case studies demonstrate its practicality: profiling-guided optimization of FlashAttention-2 to reduce interconnect stalls and synchronization overhead, and design space exploration of network-on-chip (NoC) router remapping to alleviate traffic imbalance and improve throughput.
\end{abstract}


\section{Introduction}
Recent advances in \glspl{llm} have fundamentally reshaped the design of modern accelerators.
The rapid growth in model complexity places increasing demands on parallel compute capability and on-chip memory capacity \cite{li2025largelanguagemodelinference}.
At the same time, the performance of many \gls{llm} workloads, such as attention mechanisms and fused linear algebra kernels, is increasingly dominated by fine-grained data movement and synchronization across a large number of \glspl{pe} \cite{dao2023flashattention2, kwon2023efficient}.
To efficiently support these workloads, emerging accelerator designs adopt many-core architectures with large, tightly-coupled, software-managed \gls{spm} \cite{Cavalcante2021MemPool, lcm_2024, hammerblade_2024}.
Compared to cache-based designs, \gls{spm}-centric architectures enable predictable access behavior and explicit control over data movement, thereby achieving high computational efficiency \cite{spm_asplos_2023}.

This architectural shift raises new challenges.
First, fast simulation and prototyping become increasingly difficult as many-core designs scale in both \gls{pe} count and interconnect complexity~\cite{ZARRIN2017168}.
\gls{spm}-centric architectures further enlarge the design space because software-controlled memory management interacts tightly with hardware features, making effective \gls{dse} critical for identifying optimal configurations.
Consequently, software--hardware co-design on such systems depends on end-to-end, system-level performance evaluation with timely feedback, thus motivating the need for fast and agile modeling approaches.

A second key challenge is accurately modeling fine-grained, latency-sensitive interconnect behavior in \gls{spm}-centric many-core architectures.
Even single-cycle timing variations along low-latency paths can significantly affect end-to-end performance, especially under intensive word-level accesses.
This challenge is compounded by multi-level interconnect hierarchies with heterogeneous topologies, making it essential to capture timing effects precisely and consistently across multiple scales.

This paper presents a precise and fast modeling approach for latency-sensitive, large-scale \gls{spm}-centric many-core systems.
We implement our methodology in GVSoC~\cite{gvsoc_2021}, a high-speed, event-driven simulator widely used for architectural exploration and software development, and apply it to TeraNoC~\cite{teranoc_2025}, a 1024-core architecture featuring a 4\,MiB globally shared L1 \gls{spm} and a hybrid mesh–crossbar interconnect.
The key contributions are:

\begin{itemize}
\item Developing a fast yet accurate many-core performance model that captures fine-grained, latency-sensitive access timing variations across multiple interconnect scales, spanning local crossbars and chip-level mesh \glspl{noc}, while preserving simulation speed via appropriate abstractions.
The model achieves up to $115\times$ speedup over \gls{rtl} simulation, with an error margin below 7\%.
\item Deploying the FlashAttention-2~\cite{dao2023flashattention2} kernel on TeraNoC using the proposed model, demonstrating efficient software tuning enabled by rich, fine-grained profiling.
\item Identifying interconnect bottlenecks and conducting hardware \gls{dse} with the model for improved performance under representative workloads; specifically, introducing \gls{noc} router remapping to enhance throughput and reduce latency.
\end{itemize}


\section{Related Work}
Architecture evaluation spans a wide spectrum of abstraction levels, ranging from cycle-accurate \gls{rtl} simulation to high-level analytical or functional modeling \cite{simulation_2019, noctools_2018}.

Cycle-accurate \gls{rtl} simulation provides a golden reference for timing fidelity, but suffers from prohibitively long runtimes when applied to large-scale many-core systems \cite{simulation_2019}, making it unsuitable for rapid design iteration.
\gls{fpga}-based prototyping offers higher execution speed \cite{fast_2007}, but typically imposes constraints on system complexity due to limited hardware resources \cite{mrp_2015}, particularly for scaling the number of \glspl{pe} and modeling complex memory subsystems.
Moreover, both \gls{rtl} simulation and \gls{fpga} prototyping suffer from limited flexibility and incur substantial design and integration effort \cite{simulation_2019}.
At the other end of accuracy--speed trade-off, binary translation--based simulators enable extremely fast execution and full-system simulation \cite{qemu_2005, Riedel2021}, but their highly abstracted models are typically decoupled from hardware microarchitecture, resulting in limited fidelity for performance estimation.

Event-based simulators offer a balanced trade-off between accuracy and simulation speed and are therefore widely adopted for architectural analysis and \gls{dse}.
Classic NoC simulators such as Garnet \cite{garnet_2009}, BookSim \cite{booksim_2013}, and NoXim \cite{noxim_2015} provide detailed models of router microarchitecture, routing algorithms, and arbitration policies, enabling accurate network-level evaluation across common topologies.
These approaches assume simplified endpoints and traffic injection models, which limits their ability to capture the interactions between computation and interconnect behavior in tightly coupled memory systems.

More recent work has pushed toward improved scalability and timing accuracy.
DARSIM \cite{lis2010darsim} parallelizes \gls{noc} simulation with periodic synchronization to scale to larger networks and improve throughput, while BZSim \cite{bzsim_2024} accelerates simulation of large-scale many-core systems by combining analytical latency estimation with selective detailed simulation.
ONNXim \cite{onnxim_2024} achieves cycle-accurate modeling of memory subsystems and \gls{noc} for multi-core \gls{npu} targeting \gls{dnn} inference.

Despite these advances, existing interconnect modeling approaches are primarily built around packet-level traffic and are validated in regimes where end-to-end network latency is typically on the order of tens of cycles.
In contrast, latency-sensitive interconnects, which support fine-grained communication between \glspl{pe} and multi-banked L1 \glspl{spm}, are modeled in an abstract fashion, with limited timing accuracy.

Furthermore, prior works generally lack end-to-end, system-level profiling under real workloads as well as the associated \gls{dse} support.
Our work instead targets latency-sensitive, word-level \gls{spm} interconnect modeling, and emphasizes calibrated end-to-end timing accuracy together with system-level profiling to enable iterative software tuning and hardware \gls{dse} for performance optimization. In this context, an error below 10\% relative to cycle-accurate simulation is highly desirable.

Table~\ref{tab:sim_compare} summarizes representative approaches along the criteria most relevant to this work.

\begin{table}[t]
  \centering
  \caption{Comparison of representative simulation approaches.}
  \label{tab:sim_compare}
  \footnotesize
  \setlength{\tabcolsep}{2.5pt}
  \begin{tabular}{@{}l l l l l l@{}}
    \toprule
    \textbf{Simulator} & \textbf{Method} & \textbf{Gran.} &
    \textbf{Topology} & \textbf{Error vs \gls{rtl}} & \textbf{Scope} \\
    \midrule
    RTL sim. & cycle-accurate & as-impl. & as-impl. & reference & system \\
    QEMU \cite{qemu_2005} & binary-tran. & instr. & N/A & N/A & system \\
    Garnet \cite{garnet_2009} & event-based & packet & multiple & 10\%--15\% & \gls{noc} \\
    DARSIM \cite{lis2010darsim} & traffic-trace & packet & mesh & \textless 20\% & \gls{noc} \\
    BZSim \cite{bzsim_2024} & hybrid & packet & multiple & 10\%--20\% & \gls{noc} \\
    ONNXim \cite{onnxim_2024} & event-based & packet & multiple & 5\% on \gls{noc} & system \\
    \textbf{This work} & \textbf{event-based} & \color[HTML]{009900} \textbf{word} & \color[HTML]{009900} \textbf{multiple} & \color[HTML]{009900} \textbf{\textless 7\% E2E} & \color[HTML]{009900} \textbf{system} \\
    \bottomrule
  \end{tabular}
  
  \vspace{0.2\baselineskip}
  {\scriptsize\raggedright
  \textbf{Gran.}: traffic granularity; \textbf{N/A}: not applicable; \textbf{E2E}: end-to-end.\par}
\end{table}

\section{Simulator and Testbed Architecture}
\label{sec:background}
This section provides a brief overview of the simulation framework used in this work, GVSoC, and the target architecture for demonstration, TeraNoC.

\subsection{GVSoC}
GVSoC~\cite{gvsoc_2021} is an open-source high-speed, event-driven architectural simulator designed for timing-aware, full-system simulation.
Rather than enforcing strict cycle-wise execution, GVSoC models architectural components as interacting entities whose state updates are triggered by timestamped events.
This execution model enables flexible timing abstraction while preserving causal correctness across the simulated system.

Simulation in GVSoC is driven by clock engines, each associated with an event queue.
Events are scheduled with explicit target timestamps and processed in temporal order.
When an event is triggered, its handler is executed by the corresponding clock engine to update the architectural state and therefore model the functional behavior of the component.
Based on the state, the handler may schedule subsequent events with new timestamps, thereby capturing the timing behavior of hardware operations.

Communication between components in GVSoC is expressed via abstract request calls, allowing architectural interactions to be modeled independently of any particular microarchitectural implementation.
This interface gives model developers flexibility in selecting the desired abstraction level, ranging from coarse-grained functional interactions to fine-grained, timing-aware communication.
For example, a memory access can be modeled either as an instantaneous abstracted operation, or as a multi-cycle process that incurs latency and occupies bandwidth under defined timing constraints.

\subsection{TeraNoC Architecture}
\label{subsec:teranoc_arch}

\begin{figure}[t]
  \centering
  \includegraphics[width=\linewidth]{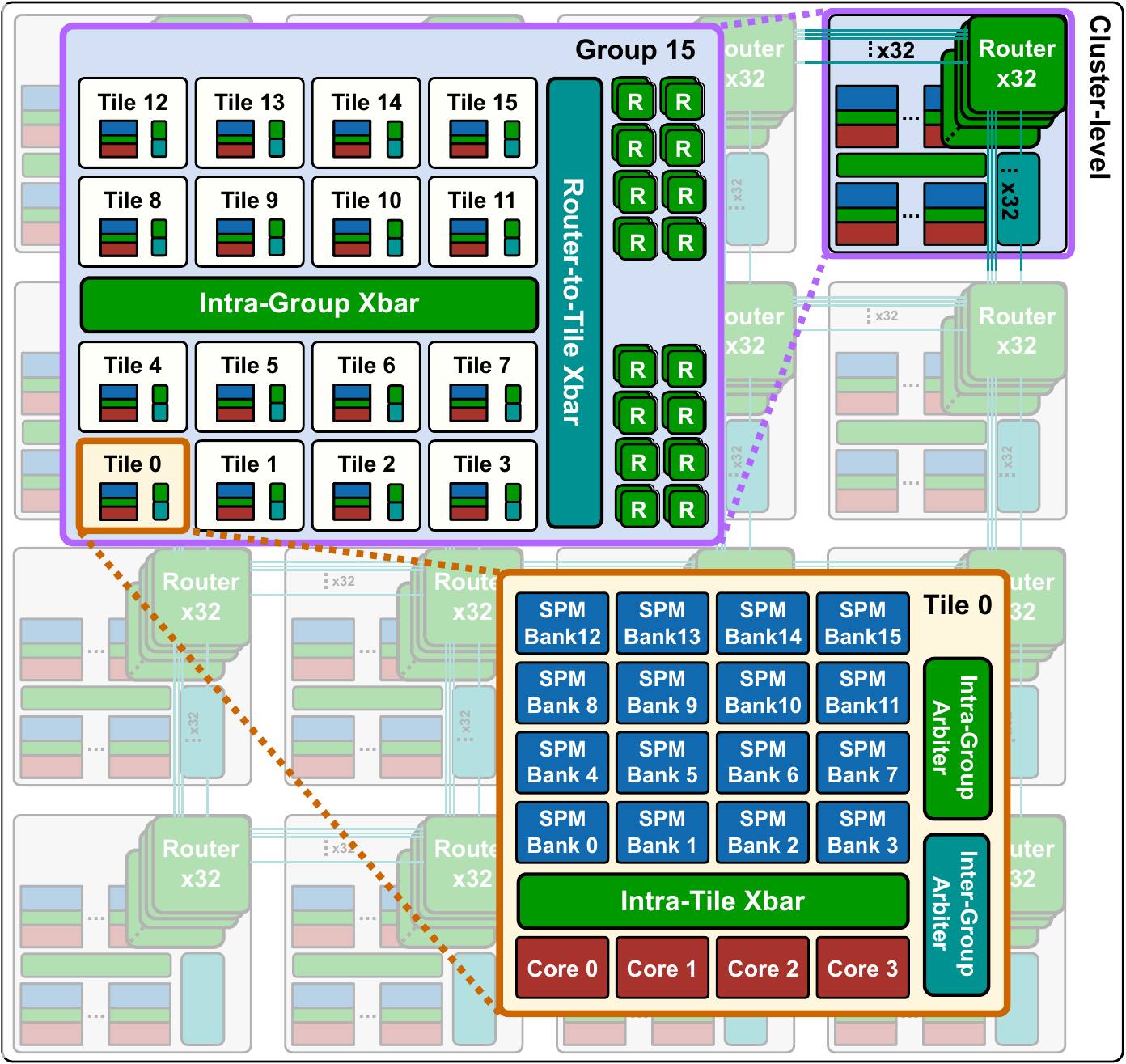}
  \caption{TeraNoC Architecture.}
  \label{fig:teranoc_arch}
\end{figure}

TeraNoC~\cite{teranoc_2025} is designed to scale a tightly coupled, shared-L1 \gls{spm} cluster beyond 1000 cores while preserving low access latency and high bandwidth by using a hybrid mesh--crossbar interconnect.
To balance single-core utilization and many-core scalability, the cluster is partitioned hierarchically, as shown in Fig.~\ref{fig:teranoc_arch}: 
a base-level block (Tile) integrates a subset of \glspl{pe} and a portion of multi-banked L1 \gls{spm}, while multiple Tiles form a higher-level block (Group). 
This organization creates distinct interconnect regimes (intra-Tile, intra-Group, and inter-Group), which are central to the timing behavior targeted by our GVSoC model.

Within a Tile, each \gls{pe} is a single-issue, single-stage, 32-bit RV32IA Snitch core, extended with an \gls{ipu} (int32/16b) and a \gls{fpu} (fp32/16b).
To tolerate memory access latency, the \gls{lsu} of each \gls{pe} supports multiple outstanding transactions.
The L1 \gls{spm} is software-managed and physically implemented as many small SRAM banks; banking and address interleaving expose high aggregate bandwidth, while bank conflicts and crossbar arbitration determine contention under load.
\glspl{pe} access the local L1 bank portion through intra-Tile crossbars to provide low-latency, high-bandwidth \gls{spm} accesses.
For requests targeting banks outside the local Tile, arbitration at the Tile boundary forwards transactions to either an intra-Group crossbar for accessing other Tiles within the same Group, or injects them into the inter-Group mesh  \gls{noc} for long-distance accesses.
On the destination side, incoming traffic is received at the Group interface, forwarded to the target Tile, and finally delivered to the addressed \gls{spm} bank through the intra-Tile crossbar.
Requests and responses traverse separate networks, allowing dataless requests and data-bearing responses to be handled independently.

At the Cluster level, Groups are interconnected by a 2D mesh \gls{noc} to obtain a regular routing pattern with low wiring overhead.
The mesh is implemented as multiple parallel, narrow 32-bit channels; the number of channels (and thus routers) per Tile is a design-time parameter that can be tuned to match available wiring resources and the target bandwidth.
In the baseline configuration used in this paper, the parallel mesh routers are statically mapped to the inter-Group ports of Tiles. 
This hybrid hierarchy enables very low latency for local and intra-Group accesses (on the order of a few cycles), while keeping remote-Group accesses bounded to modest multi-hop delays (e.g., 7 cycles between adjacent Groups and 13.7 cycles on average in the evaluated $4\times4$ mesh configuration).

\begin{table}[t]
  \centering
  \footnotesize
  \caption{TeraNoC configuration used in evaluation.}
  \label{tab:teranoc_config}
  \begin{tabularx}{\columnwidth}{l l X}
    \toprule
    \textbf{Level} & \textbf{Item} & \textbf{Baseline configuration} \\
    \midrule
    Tile
    & \glspl{pe} & 4 Snitch Cores \\
    & L1 \gls{spm} & 16 banks $\times$ 1\,KiB \\
    & Ports & 1 intra-Group + 2 inter-Group Ports \\
    \midrule
    Group
    & Tiles & 16 Tiles \\
    & Routers & 32 Routers (2 per Tile) \\
    \midrule
    Cluster
    & Groups & 16 Groups \\
    & Topology & $4{\times}4$ 2D mesh \\
    & \gls{noc} instances & 32 request + 32 response \glspl{noc} \\
    \bottomrule
  \end{tabularx}
\end{table}

Table~\ref{tab:teranoc_config} summarizes the baseline TeraNoC configuration used in our evaluation.
While we report results for this concrete setup, the GVSoC model exposes all listed quantities as configurable parameters to enable systematic \gls{dse}.

\section{Modeling \& Evaluation}
\label{sec:modeling}
This section presents our modeling methodology and its end-to-end validation on the TeraNoC architecture.

\subsection{Request Propagation and Timing Modeling}
Building on the request-based component interface provided by GVSoC, we model \gls{spm} interconnect behavior as the propagation of abstract requests among architectural components.
We focus on the functional semantics of data movement and the accumulation of timing effects along each request’s path, while abstracting away other implementation-specific details.

Each request represents a memory-mapped operation issued by an active component such as a \gls{pe} or a \gls{dma} engine.
It encapsulates the information required for interconnect modeling, including the target address, operation type, payload, and timing metadata.
Requests are processed and forwarded through a sequence of components that form the interconnect topology.
At each hop, the component processes the request using its local state and the request's attributes, then dispatches it to the next hop or generates a response if the destination is reached.

Algorithm~\ref{alg:req_proc} gives an overview of this process.
For each hop, the routing decision ($p$) is determined by the request's target address and the component's routing policy, while arbitration and congestion effects are evaluated using the request's timing metadata and the component's internal state.
In our model, the timing metadata is mainly represented by two complementary terms: \emph{latency} ($L$), which accumulates the component’s intrinsic delay and any additional waiting due to arbitration or congestion, and \emph{duration} ($D$), which captures the payload serialization determined by the component's bandwidth.
Accordingly, the component updates the timing metadata and its internal state before dispatching the request onward.
The propagation terminates at the destination component, which serves the operation and returns the response to the initiator.

\begin{algorithm}[t]
  \caption{Per-hop request processing.}
  \label{alg:req_proc}
  \small
  \begin{algorithmic}[1]
    \Statex \textbf{Processing of request $r$ at component $c$}
    \Statex \textit{// Routing decision}
    \State $p \gets c.\mathrm{Route}(r.addr)$
    \Statex \textit{// Timing update}
    \State $L \gets c.\mathrm{BaseLatency}(p, r) + c.\mathrm{ArbitrationDelay}(p, r)$
    \State $D \gets c.\mathrm{TransferDuration}(p, r)$
    \State $r.latency \gets r.latency + L$
    \State $r.duration \gets \max(r.duration,\; D)$
    \Statex \textit{// State update and dispatch}
    \State $c.\mathrm{UpdateState}(p, r)$
    \State $c.\mathrm{Dispatch}(p, r)$
  \end{algorithmic}
\end{algorithm}

Through this propagation process, end-to-end communication timing emerges naturally as the accumulation of per-component updates, rather than being imposed as an analytically derived delay.
This approach captures fine-grained latency variations arising from contention and arbitration, which are particularly critical for latency-sensitive \gls{spm} accesses.
At the same time, this request-based abstraction supports modeling traffic at various granularities, from word-level accesses to coarse-grained bulk transfers, within a unified framework.

\subsection{Router-Based Interconnect Abstraction}
We propose a generic router template as a modeling primitive for interconnects.
The router serves as an abstract, all-to-all switchbox with user-defined behavior, making it a common building block across interconnect layers.
A router instance can be configured with a variable number of input and output ports, routing and arbitration policies, as well as timing parameters such as bandwidth and latency, allowing flexible construction of diverse interconnect topologies.

As an example, the intra-Group crossbar can be modeled using a router whose input and output ports correspond to the number of Tiles, with a bandwidth of 4~bytes/cycle and a zero-load latency of one cycle.
The router implements fully-interleaved address mapping for target selection and uses a round-robin arbitration policy to resolve contention.
By instantiating the same router primitive with different architectural descriptions, the model uniformly represents both the intra-Group multi-level crossbar and the inter-Group mesh \gls{noc}.

To keep the model compact and reduce simulation overhead, we avoid a one-to-one mapping between hardware blocks and model components by folding simple interconnect elements (e.g., spill registers) into adjacent router components.
This approximation introduces negligible timing error while reducing the number of components and request propagation steps.

\subsection{Timing Accuracy and Performance Trade-offs}
GVSoC supports both synchronous and asynchronous execution strategies for modeling request propagation through the interconnect.
In synchronous execution, a request is fully processed within a single event handler; multi-hop propagation across components is realized via nested request calls.
This approach achieves high simulation speed, but provides limited visibility into concurrent requests, which restricts the fidelity of contention-related timing effects.
In contrast, asynchronous execution allows components to buffer requests and forward them in subsequent simulation events, enabling more hardware-like modeling of arbitration, contention, and hence latency variations.
The improved timing fidelity comes at higher simulation overhead compared with synchronous execution, as each request may span multiple events.

In our model, we selectively apply these two strategies across the interconnect hierarchy based on contention characteristics and their impact on end-to-end performance.
For intra-Group communication, which is realized by a multi-level crossbar hierarchy, we model the request path asynchronously to capture fine-grained contention and arbitration under highly concurrent, word-level accesses.
In contrast, the response path within the same crossbar hierarchy typically exhibits much lower contention under typical workloads; we therefore model responses synchronously with lightweight congestion handling to reduce simulation overhead while preserving timing fidelity in the dominant contention scenarios.
For the inter-Group mesh \gls{noc}, both request and response paths are modeled asynchronously.
Unlike the intra-Group crossbar, mesh traffic traverses multiple hops over shared links, and response traffic can experience non-negligible contention comparable to requests.
Overall, this selective combination achieves a favorable trade-off between timing accuracy and simulation performance: contention-sensitive paths use higher-fidelity asynchronous modeling, while less critical paths are simplified without compromising end-to-end performance evaluation.

\subsection{Profiling Support}
Our model exposes fine-grained profiling data across \glspl{pe}, interconnect components, and memories, enabling insightful bottleneck analysis for both software tuning and hardware design exploration.
Table~\ref{tab:profiling_support} summarizes the profiling capabilities supported by our framework.

\begin{table}[t]
  \centering
  \caption{Supported profiling outputs.}
  \label{tab:profiling_support}
  \setlength{\tabcolsep}{4pt}
  \footnotesize
  \begin{tabularx}{\columnwidth}{l l X}
    \toprule
    \textbf{Scope} & \textbf{Output} & \textbf{Collected information} \\
    \midrule
    \gls{pe}
    & Trace & Instruction execution trace; per-\gls{pe} state. \\
    & Perf.\ counter & Inst/load/store counts; stall-cycle breakdown. \\
    \midrule
    Interconnect
    & Trace & Transfer events: enqueue, arbitrate, dispatch. \\
    & Utilization & Link/port busy cycles; transfer event counts. \\
    & Congestion & Queue occ.; port conflicts and blocking. \\
    \midrule
    Memory
    & Trace & Memory access log. \\
    & Conflict & Bank conflict statistics. \\
    \bottomrule
  \end{tabularx}
\end{table}

\subsection{Benchmark Results}
To evaluate the accuracy and efficiency of the proposed GVSoC model, we select a diverse set of workloads spanning traditional linear algebra and signal processing, as well as modern machine learning kernels.

\begin{figure*}[t]
  \centering
  \includegraphics[width=\textwidth]{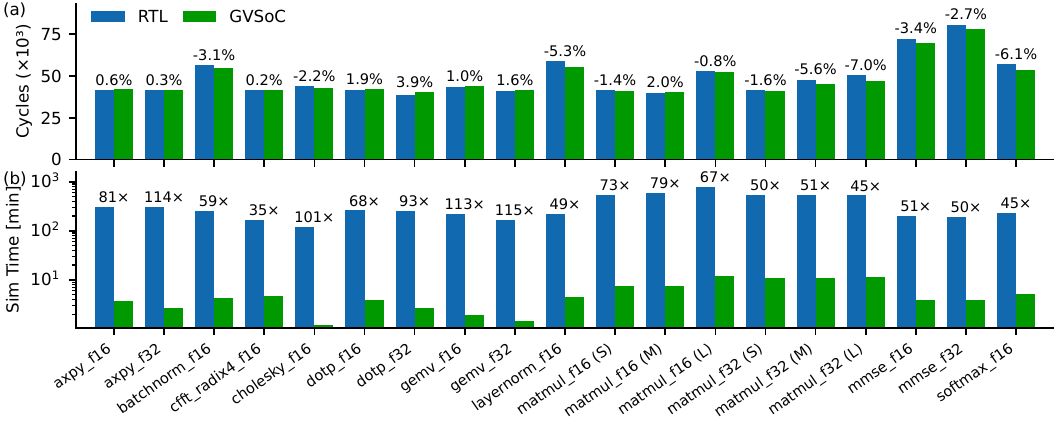}
  \caption{Accuracy and simulation runtime of the proposed GVSoC model across evaluated workloads.}
  \label{fig:accuracy_speedup}
\end{figure*}

Fig.~\ref{fig:accuracy_speedup} presents the accuracy and efficiency of the GVSoC model across all evaluated kernels.
Fig.~\ref{fig:accuracy_speedup}(a) compares GVSoC results against cycle-accurate \gls{rtl} simulation, showing that the GVSoC model closely tracks \gls{rtl} behavior with relative error consistently below 7\% across all workloads.
Fig.~\ref{fig:accuracy_speedup}(b) reports the simulation runtime, demonstrating substantial acceleration over \gls{rtl} simulation.
Depending on the workload, the GVSoC model achieves speedups of more than two orders of magnitude, enabling rapid design iteration.

\section{End-to-end Software Deployment}
\label{sec:sw_deploy}
The proposed model enables an end-to-end software development workflow on TeraNoC by combining fast simulation with actionable profiling feedback.
To demonstrate its practical value, we deploy FlashAttention-2 and present a profiling-guided optimization case study, which would be difficult to achieve with cycle-accurate \gls{rtl} simulation at scale.

FlashAttention-2~\cite{dao2023flashattention2} adopts a tiled formulation of self-attention, where the query matrix is partitioned along the sequence dimension and the attention score computation is further decomposed into smaller tiles along the key/value dimension.
This blocking strategy avoids materializing the full attention matrix and enables streaming execution with bounded on-chip storage.
In this work, we follow the deployment and execution methodology described in~\cite{das_2025} to implement the FlashAttention-2 kernel.

\begin{figure}[t]
  \centering
  \includegraphics[width=\columnwidth]{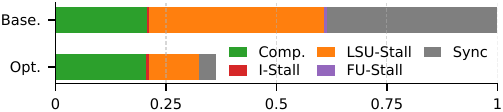}
  \caption{Execution breakdown (normalized) of FlashAttention-2.}
  \label{fig:sw_exec_breakdown}
\end{figure}

Leveraging the model's profiling capabilities (Table~\ref{tab:profiling_support}), the execution breakdown in Fig.~\ref{fig:sw_exec_breakdown}~(Base.) reveals low \gls{pe} utilization with substantial interconnect-induced stalls, indicating the dominance of long-latency data accesses.
The high synchronization overhead further suggests execution imbalance across \glspl{pe}.
Interconnect profiling attributes this imbalance to concentrated port conflicts at the remote request input ports of a small subset of Tiles, especially at the beginning of each inner tiling loop. This creates non-uniform service latency, which in turn causes progress divergence across \glspl{pe}.

Guided by these observations, we apply two optimizations: (i) reduce long-range traffic over inter-Group \glspl{noc} to lower round-trip latency, and (ii) smooth concurrent accesses to shared data blocks to mitigate Tile-level port conflicts.
Within the existing tiling scheme, we reorganize the workload distribution across \glspl{pe} so that query accesses in score computation and all accesses in online softmax remain within the local Group, effectively reducing long-latency \gls{spm} accesses.
Moreover, we introduce staggered start offsets across \glspl{pe}, thereby smoothing the temporal access pattern and reducing peak contention at Tile-level ports.
Fig.~\ref{fig:sw_exec_breakdown}~(Opt.) shows that these changes substantially reduce interconnect-induced stalls and synchronization overhead, improving overall efficiency.
Overall, this case study highlights the model’s essential role in end-to-end software deployment.
By enabling rapid iteration and providing insightful profiling that correlates \gls{pe} execution breakdown with interconnect congestion statistics, the model guides targeted optimizations.

\section{Hardware Design Exploration}
\label{sec:hw_dse}
The proposed model enables rapid, profiling-driven hardware design exploration for interconnect-centric systems.
By exposing fine-grained performance metrics and supporting fast what-if evaluations, it allows researchers to identify architectural bottlenecks under interconnect-intensive workloads and to iterate on design alternatives efficiently.
Using this framework, we present a representative \gls{dse} case study on TeraNoC that explores \gls{noc} router remapping to improve throughput and reduce latency by alleviating traffic imbalance.

When profiling interconnect-intensive workloads with frequent global memory accesses, such as \gls{matmul}, we observe a substantial number of \gls{pe} stalls caused by congestion in the mesh \gls{noc} despite modest overall link utilization, indicating inefficient traffic distribution.
Using link utilization and congestion statistics, we further identify two forms of imbalance: (i) \emph{spatial} imbalance, where utilization differs significantly across parallel meshes within the same interval, and (ii) \emph{temporal} imbalance, where congestion hotspots shift over time within a mesh as access patterns evolve.
These results suggest that a considerable fraction of interconnect capacity remains underutilized, motivating mechanisms that better exploit idle links to improve overall throughput.

\begin{figure}[t]
  \centering
  \includegraphics[width=\columnwidth]{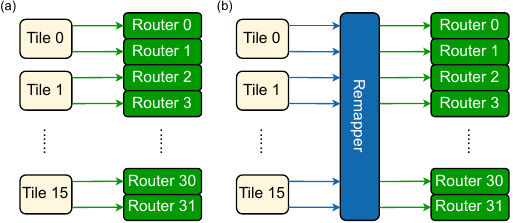}
  \caption{Router remapping architecture.}
  \label{fig:router_remapping_arch}
\end{figure}

As discussed in Section~\ref{subsec:teranoc_arch}, TeraNoC employs multiple parallel mesh \glspl{noc} for inter-Group traffic, which are statically mapped to the inter-Group ports of Tiles.
Fig.~\ref{fig:router_remapping_arch}~(a) illustrates the baseline configuration, where each Tile is connected to two fixed \gls{noc} routers.
This static connection prevents traffic injected by a Tile from being redistributed across \gls{noc} instances under unbalanced load, limiting the ability to leverage idle network capacity.
To address this limitation, we introduce router remapping by inserting a crossbar-like switch between Tiles and \gls{noc} routers, which dynamically maps Tile ports to \gls{noc} routers, as shown in Fig.~\ref{fig:router_remapping_arch}~(b).
The mapping is updated every cycle using a deterministic pseudo-random schedule, which is hardware-friendly and achieves performance close to true-random remapping in our simulations.
Using our model, we observe an overall performance improvement of approximately 10\%, while interconnect profiling further confirms a more balanced utilization across network links.

\begin{figure}[t]
  \centering
  \includegraphics[width=\columnwidth]{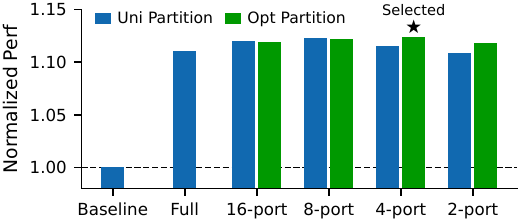}
  \caption{\gls{dse} of router remapping configurations.}
  \label{fig:router_remapping_dse}
\end{figure}

From a physical implementation perspective, the routing complexity of remapper grows quadratically with the number of ports, making a monolithic 32-to-32 full remapper impractical.
To address this limitation, we decompose the remapping functionality into multiple smaller remappers with limited port counts, thereby partitioning the remapping space into independent regions, and use our framework to explore the trade-off between remapper granularity and performance.
We use the optimized \gls{matmul} kernel (Section~\ref{sec:sw_deploy}) as the representative workload for this study.
As shown in Fig.~\ref{fig:router_remapping_dse}, the eight-port remapper delivers the best performance, while the four-port remapper is close to this optimum.

Further profiling reveals that \gls{spm} accesses from Tiles with adjacent identifiers exhibit similar patterns within a given time window, whereas logically distant Tiles show more divergent traffic behavior.
Exploiting this observation, we further adjust the remapper partitioning strategy to better align with traffic locality, yielding additional gains over uniform partitioning.
Combining four-port remappers with optimized partitioning achieves the best overall performance and is therefore selected as the final configuration.
The magnitude of improvement varies across kernels because their traffic patterns differ: kernels with bursty and spatially skewed access patterns benefit the most from remapping, as it mitigates persistent hotspots and increases effective bandwidth.
Table~\ref{tab:router_remapping_perf} summarizes the resulting performance improvements across all evaluated global interconnect–intensive workloads.

\begin{table}[t]
  \centering
  \caption{Performance improvement of \gls{noc} router remapping.}
  \label{tab:router_remapping_perf}
  \small
  \setlength{\tabcolsep}{5.5pt}
  \begin{tabular}{lcccccc}
    \hline
    \textbf{Kernel} & BN & Chol. & LN & GEMM & MMSE & SM \\
    \hline
    \textbf{Impr.} & 11.5\% & 12.2\% & 47.1\% & 12.3\% & 9.5\% & 33.8\% \\
    \hline
  \end{tabular}
  
  \vspace{2pt}
  {\footnotesize\raggedright
  BN, LN, and SM denote BatchNorm, LayerNorm, and Softmax, respectively.\par
  }
\end{table}

\section{Conclusion}
\label{sec:conclusion}
In this work, we presented an accurate and efficient end-to-end modeling approach for tightly coupled, \gls{spm}-centric many-core architectures, and applied it to TeraNoC as a representative target.
By capturing the micro-level timing behaviors of fine-grained, latency-sensitive \gls{spm} accesses while abstracting unnecessary hardware details, the model achieves up to 115$\times$ speedup over \gls{rtl} simulation and delivers very accurate performance estimates within 7\% of cycle-accurate results.
The model also provides detailed profiling insights across \glspl{pe} and the interconnect, substantially strengthening performance analysis for hardware and software development and optimization.
We demonstrated the practical use of the framework through end-to-end deployment and optimization of FlashAttention-2, where profiling-guided tuning of \gls{spm} accesses exposes and mitigates communication bottlenecks.
Finally, we demonstrated hardware \gls{dse} via a \gls{noc} router remapping case study that alleviates traffic imbalance and improves throughput for interconnect-intensive kernels.
Together, these results highlight the model’s usefulness for software--hardware co-design and \gls{dse} for performance optimization.

\ifblind
\else
\section*{Acknowledgment}
This work has received funding from the Swiss State Secretariat for Education, Research, and Innovation (SERI) under the SwissChips initiative.
\fi

\bibliographystyle{IEEEtran}
\bibliography{bibliography/bibliography}

\end{document}

%% file: settings/packages.tex
\usepackage{cite}
\usepackage{amsmath,amssymb,amsfonts}
\usepackage{algorithm}
\usepackage{algpseudocode}
\usepackage{graphicx}
\usepackage{textcomp}
\usepackage{xcolor}
\usepackage[acronym]{glossaries-extra}
\usepackage{booktabs}
\usepackage{tabularx}
\usepackage[hidelinks]{hyperref}

%% file: settings/acronyms.tex
\setabbreviationstyle[acronym]{long-short}

\newacronym{llm}{LLM}{Large Language Model}
\newacronym{pe}{PE}{Processing Element}
\newacronym{spm}{SPM}{Scratchpad Memory}
\newacronym{rtl}{RTL}{Register-Transfer Level}
\newacronym{fpga}{FPGA}{Field-Programmable Gate Array}
\newacronym{dse}{DSE}{Design Space Exploration}
\newacronym{noc}{NoC}{Network-on-Chip}
\newacronym{npu}{NPU}{Neural Processing Unit}
\newacronym{dnn}{DNN}{Deep Neural Network}
\newacronym{ipu}{IPU}{Integer Processing Unit}
\newacronym{fpu}{FPU}{Floating-Point Unit}
\newacronym{dma}{DMA}{Direct Memory Access}
\newacronym{matmul}{MatMul}{Matrix Multiplication}
\newacronym{lsu}{LSU}{Load Store Unit}

%% file: settings/authorship.tex
\ifblind
\author{Anonymous Authors}
\else
\author{
  \IEEEauthorblockN{
    Yinrong Li\IEEEauthorrefmark{1},
    Zexin Fu\IEEEauthorrefmark{1},
    Yichao Zhang\IEEEauthorrefmark{1},
    Germain Haugou\IEEEauthorrefmark{1},
    Chi Zhang\IEEEauthorrefmark{1}, \\
    Marco Bertuletti\IEEEauthorrefmark{1},
    Bowen Wang\IEEEauthorrefmark{1},
    Luca Benini\IEEEauthorrefmark{1}\IEEEauthorrefmark{2}
  }
  \IEEEauthorblockA{
    \textit{\IEEEauthorrefmark{1}Integrated Systems Laboratory (IIS), ETH Zurich, Zurich, Switzerland} \\
    \textit{\IEEEauthorrefmark{2}Department of Electrical, Electronic, and Information Engineering (DEI), University of Bologna, Bologna, Italy} \\
  yinrli@student.ethz.ch, \{zexifu, yiczhang, haugoug, chizhang, mbertuletti, bowwang, lbenini\}@iis.ee.ethz.ch
  }
}
\fi

%% file: bibliography/bibliography.bib
@misc{li2025largelanguagemodelinference,
  author = {Li, Jinhao and Xu, Jiaming and Huang, Shan and Chen, Yonghua and Li, Wen and Liu, Jun and Lian, Yaoxiu and Pan, Jiayi and Ding, Li and Zhou, Hao and Wang, Yu and Dai, Guohao},
  title = {Large Language Model Inference Acceleration: A Comprehensive Hardware Perspective},
  year = {2025},
  note = {arXiv:2410.04466}
}

@inproceedings{dao2023flashattention2,
  author = {Dao, Tri},
  title = {Flash{A}ttention-2: Faster Attention with Better Parallelism and Work Partitioning},
  booktitle = {ICLR},
  year = {2024}
}

@inproceedings{kwon2023efficient,
  author = {Kwon, Woosuk and Li, Zhuohan and Zhuang, Siyuan and Sheng, Ying and Zheng, Lianmin and Yu, Cody Hao and Gonzalez, Joseph E. and Zhang, Hao and Stoica, Ion},
  title = {Efficient Memory Management for {LLM} Serving with {PagedAttention}},
  booktitle = {SOSP},
  year = {2023},
  pages = {611--626}
}

@inproceedings{Cavalcante2021MemPool,
  author = {Cavalcante, Matheus and Riedel, Samuel and Pullini, Antonio and Benini, Luca},
  title = {{MemPool}: A Shared-{L1} Memory Many-Core Cluster with a Low-Latency Interconnect},
  booktitle = {DATE},
  year = {2021},
  pages = {701--706}
}

@inproceedings{hammerblade_2024,
  author = {Jung, Dai Cheol and Ruttenberg, Max and Gao, Paul and Davidson, Scott and Petrisko, Daniel and Li, Kangli and Kamath, Aditya K and Cheng, Lin and Xie, Shaolin and Pan, Peitian and Zhao, Zhongyuan and Yue, Zichao and Veluri, Bandhav and Muralitharan, Sripathi and Sampson, Adrian and Lumsdaine, Andrew and Zhang, Zhiru and Batten, Christopher and Oskin, Mark and Richmond, Dustin and Taylor, Michael Bedford},
  title = {Scalable, Programmable and Dense: The {HammerBlade} Open-Source {RISC-V} Manycore},
  booktitle = {ISCA},
  year = {2024},
  pages = {770--784}
}

@inproceedings{spm_asplos_2023,
  author = {Cheng, Lin and Ruttenberg, Max and Jung, Dai Cheol and Richmond, Dustin and Taylor, Michael and Oskin, Mark and Batten, Christopher},
  title = {Supporting Dynamic Task Parallelism on Manycore Architectures with Software-Managed Scratchpad Memories},
  booktitle = {ASPLOS},
  year = {2023},
  pages = {46--58}
}

@inproceedings{lcm_2024,
  author = {Lai, Chengtao and Zhou, Zhongchun and Poptani, Akash and Zhang, Wei},
  title = {{LCM}: {LLM}-focused Hybrid {SPM}-cache Architecture with Cache Management for Multi-Core {AI} Accelerators},
  booktitle = {ICS},
  year = {2024},
  pages = {62--73}
}

@article{ZARRIN2017168,
  author = {Zarrin, Javad and Aguiar, Rui L. and Barraca, João Paulo},
  title = {Manycore simulation for peta-scale system design: Motivation, tools, challenges and prospects},
  journal = {Simul. Model. Pract. Theory},
  volume = {72},
  year = {2017},
  pages = {168--201}
}

@article{simulation_2019,
  author = {Akram, Ayaz and Sawalha, Lina},
  title = {A Survey of Computer Architecture Simulation Techniques and Tools},
  journal = {IEEE Access},
  volume = {7},
  year = {2019},
  pages = {78120--78145}
}

@article{noctools_2018,
  author = {Khan, Sarzamin and Anjum, Sheraz and Gulzari, Usman Ali and Torres, Frank Sill},
  title = {Comparative analysis of network-on-chip simulation tools},
  journal = {IET Computers \& Digital Techniques},
  volume = {12},
  pages = {30-38},
  year = {2018}
}

@inproceedings{fast_2007,
  author = {Chiou, Derek and Sunwoo, Dam and Kim, Joonsoo and Patil, Nikhil A. and Reinhart, William and Johnson, Darrel Eric and Keefe, Jebediah and Angepat, Hari},
  title = {{FPGA}-Accelerated Simulation Technologies ({FAST}): Fast, Full-System, Cycle-Accurate Simulators},
  booktitle = {MICRO},
  year = {2007},
  pages = {249--261}
}

@inproceedings{mrp_2015,
  author = {Chen, Xinke and Zhang, Guangfei and Wang, Huandong and Wu, Ruiyang and Wu, Peng and Zhang, Longbing},
  title = {{MRP}: Mix real cores and pseudo cores for {FPGA}-based chip-multiprocessor simulation},
  booktitle = {DATE},
  year = {2015},
  pages = {211--216}
}

@inproceedings{qemu_2005,
  author = {Bellard, Fabrice},
  title = {{QEMU}, a fast and portable dynamic translator},
  booktitle = {USENIX ATC},
  year = {2005},
  pages = {41}
}

@inproceedings{Riedel2021,
  author = {Riedel, Samuel and Schuiki, Fabian and Scheffler, Paul and Zaruba, Florian and Benini, Luca},
  title = {Banshee: A Fast {LLVM}-Based {RISC-V} Binary Translator},
  booktitle = {ICCAD},
  year = {2021},
  pages = {1105--1113}
}

@inproceedings{garnet_2009,
  author = {Agarwal, Niket and Krishna, Tushar and Peh, Li-Shiuan and Jha, Niraj K.},
  title = {{GARNET}: A detailed on-chip network model inside a full-system simulator},
  booktitle = {ISPASS},
  year = {2009},
  pages = {33--42}
}

@inproceedings{booksim_2013,
  author = {Jiang, Nan and Becker, Daniel U. and Michelogiannakis, George and Balfour, James and Towles, Brian and Shaw, D. E. and Kim, John and Dally, William J.},
  title = {A detailed and flexible cycle-accurate Network-on-Chip simulator},
  booktitle = {ISPASS},
  year = {2013},
  pages = {86--96}
}

@inproceedings{noxim_2015,
  author = {Catania, Vincenzo and Mineo, Andrea and Monteleone, Salvatore and Palesi, Maurizio and Patti, Davide},
  title = {Noxim: An open, extensible and cycle-accurate network on chip simulator},
  booktitle = {ASAP},
  year = {2015},
  pages = {162--163}
}

@inproceedings{lis2010darsim,
  author={Lis, Mieszko and Shim, Keun Sup and Cho, Myong Hyon and Ren, Pengju and Khan, Omer and Devadas, Srinivas},
  title={DARSIM: a parallel cycle-level NoC simulator},
  booktitle = {MoBS},
  year={2010}
}

@article{onnxim_2024,
  author = {Ham, Hyungkyu and Yang, Wonhyuk and Shin, Yunseon and Woo, Okkyun and Heo, Guseul and Lee, Sangyeop and Park, Jongse and Kim, Gwangsun},
  title = {{ONNXim}: A Fast, Cycle-Level Multi-Core {NPU} Simulator},
  journal = {IEEE Comput. Archit. Lett.},
  volume = {23},
  year = {2024},
  pages = {219--222}
}

@inproceedings{bzsim_2024,
  author = {Strikos, Panagiotis and Ejaz, Ahsen and Sourdis, Ioannis},
  title = {{BZSim}: Fast, Large-Scale Microarchitectural Simulation with Detailed Interconnect Modeling},
  booktitle = {ISPASS},
  year = {2024},
  pages = {167--178}
}

@inproceedings{teranoc_2025,
  author = {Zhang, Yichao and Fu, Zexin and Fischer, Tim and Li, Yinrong and Bertuletti, Marco and Benini, Luca},
  title = {{TeraNOC}: A Multi-Channel 32-Bit Fine-Grained, Hybrid Mesh-Crossbar {NoC} for Efficient Scale-Up},
  booktitle = {ICCD},
  year = {2025},
  pages = {610--617}
}

@inproceedings{gvsoc_2021,
  author = {Bruschi, Nazareno and Haugou, Germain and Tagliavini, Giuseppe and Conti, Francesco and Benini, Luca and Rossi, Davide},
  title = {{GVSoC}: A Highly Configurable, Fast and Accurate Full-Platform Simulator for {RISC-V} {IoT} Processors},
  booktitle = {ICCD},
  year = {2021},
  pages = {409--416}
}

@inproceedings{das_2025,
  author = {Wang, Bowen and Bertuletti, Marco and Zhang, Yichao and Jung, Victor J.B. and Benini, Luca},
  title = {A Dynamic Allocation Scheme for Adaptive Shared-Memory Mapping on Kilo-Core {RV} Clusters},
  booktitle = {ASAP},
  year = {2025},
  pages = {9--16}
}
